\documentclass{ws-procs975x65_MG12}

\usepackage[english]{babel}
\usepackage{ucs}
\usepackage[utf8x]{inputenc}
\usepackage{amsmath,amssymb}
\usepackage{times}
\usepackage[small,compact]{titlesec}

\newcommand{\R}{\ensuremath{\mathbb R}}
\newcommand{\So}{\ensuremath{\mathbb S^1}}
\newcommand{\St}{\ensuremath{\mathbb S^2}}
\newcommand{\SoXSt}{\ensuremath{\mathbb S^1\times\mathbb S^2}}

\newcommand{\rcite}[1]{Ref.~\refcite{#1}}

\graphicspath{{figures/}}

\begin{document}

\title{THE COSMIC NO-HAIR CONJECTURE\\ A STUDY OF THE NARIAI SOLUTIONS}

\author{FLORIAN BEYER}

\address{Department of Mathematics and Statistics, University of Otago,\\
P.O. Box 56, Dunedin 9054, New Zealand\\
E-mail: fbeyer@maths.otago.ac.nz}

\begin{abstract}
  In this talk, we investigate the cosmic no-hair conjecture for
  perturbed Nariai solutions within the class of Gowdy symmetric
  solutions of Einstein's field equations in vacuum with a positive
  cosmological constant. In particular, we are interested whether
  these perturbations allow to construct new cosmological black hole
  solutions.
\end{abstract}


\bodymatter

\section*{Introduction and motivation}
The successful interpretation of cosmological observations requires a
deep and fundamental understanding of rapidly expanding solutions of
Einstein's field equations beyond spatial homogeneity and isotropy.
The so-called standard model of cosmology, which is based on these
symmetry assumptions, is surprisingly consistent with current
observations\cite{Sanchez06,Spergel06}.  A keystone of the standard
model is inflation, 
i.e.\ a period characterized by accelerated expansion. 
Since one can expect that strong
inhomogeneities caused by quantum fluctuations and other physical
processes are dominant in the very early universe, it is, however,
important to study the inhomogeneous case.  A particular concern is
whether inflation is able to homogenize and isotropize \textit{generic} initial
inhomogeneities. If so, this would yield a natural explanation for the
apparent homogeneity and isotropy of our present universe, without
making ad-hoc assumptions on the initial conditions.  The conjecture
that generic expanding solutions of Einstein's field equations in
vacuum with cosmological constant $\Lambda>0$,
\begin{equation}
  \label{eq:EFE}
  G_{\mu\nu}+\Lambda g_{\mu\nu}=0,
\end{equation}
approach a particular homogeneous and isotropic solution
asymptotically, namely the de-Sitter solution, is known as the cosmic
no-hair conjecture\cite{gibbons77,Hawking82}. In the following
discussion, we use this conjecture in the form of Conjecture~1 in
\rcite{beyer09:Nariai1}.  Recall that $\Lambda>0$ in Einstein's field
equations is the simplest model for dark energy which is consistent
with all current observations; see the references above.

Although there is some support for this conjecture in special
situations\cite{Wald83,weber84,barrow88,Moniz,Kitada,Ringstrom06c},
the general case remains unclear due to the complexity of Einstein's
field equations. From this point of view, a particularly interesting
family of solutions of \eref{eq:EFE} is the class of Nariai solutions
\cite{Nariai50,Nariai51}. These are \textit{simple} solutions which
are \textit{not} consistent with the cosmic no-hair picture.  The
(standard) Nariai solution of \eref{eq:EFE} is
\begin{equation}
  \label{eq:StandardNariai}
  g=(
    -dt^2+\cosh^2t\,d\rho^2+g_{\St})/\Lambda,
\end{equation}
on the manifold $M=\R\times(\SoXSt)$ for any $\Lambda>0$.  Here,
$g_{\St}=d\theta^2+\sin^2\theta d\phi^2$ in standard polar coordinates
$(\theta,\phi)$ on $\St$. In \rcite{beyer09:Nariai1}, we also define
\textit{generalized} Nariai solutions whose properties are listed in
the references above. In the following we often speak of \textit{the}
Nariai solution when we mean any generalized Nariai solution.

It is easy to see from \eref{eq:StandardNariai} that the asymptotics
for $|t|\rightarrow\infty$ are particularly peculiar.  While the
\So-factor of the spatial slices expands exponentially for increasing
positive $t$, the volume of the \St-factor stays constant. Thus, the
expansion of this solution is anisotropic in the sense that the shear
tensor of the $t=const$-slices never approaches zero. This implies
that the cosmic no-hair picture does not hold with respect to the
foliation of $t=const$-surfaces. This alone does not mean that the
cosmic no-hair conjecture is false. On the one hand, it remains
possible that there exist other foliations of the Nariai solutions,
presumably based on non-symmetric surfaces, for which the cosmic
no-hair picture is attained. However, in \rcite{beyer09:Nariai1}, we
prove the following statement based on the results in
\rcite{Ringstrom06c}: ``The Nariai solutions do not have even a patch
of a smooth conformal boundary. The same is true for their universal
cover.''  This suggests that the Nariai solutions violate the cosmic
no-hair picture for any choice of foliation because, if a given
foliation approached a homogeneous and isotropic foliation of the
de-Sitter solution, it would presumably approach a patch of a smooth
conformal boundary. 

\section*{Perturbations and cosmological black 
hole solutions}
Hence, if the cosmic no-hair conjecture is true, the asymptotics of
the Nariai solutions should be special and should be instable under
perturbations. We expect that generic perturbations either collapse
and form a singularity in a given time direction, or if there is
expansion, they approach the de-Sitter solution.  In general, when we
speak of a perturbation of a Nariai solution in the following, we mean
a cosmological solution of the fully non-linear Einstein's field
equations \eqref{eq:EFE} whose data, on some Cauchy surface, is close
to the data on a Cauchy surface of a given Nariai solution. In
principle, we are interested in generic perturbations without
symmetries. In practical investigations, however, we must make
simplifying assumptions. In a first step in \rcite{beyer09:Nariai1},
we investigated the spatially homogeneous (but anisotropic) case of
perturbations and proceeded with the Gowdy
case\cite{Gowdy73,chrusciel1990} in \rcite{beyer09:Nariai2}.

Spatially homogeneous (but anisotropic) solutions with spatial
\SoXSt-topology of the vacuum field equations with $\Lambda>0$ are
either Schwarzschild-de-Sitter or generalized Nariai solutions
locally. This is derived in \rcite{beyer09:Nariai1} and references
therein, and yields a full description of the expected instability of
the Nariai solutions with respect to spatially homogeneous
perturbations as follows. It turns out that there is a parameter
$H^{(0)}_*$ introduced in \rcite{beyer09:Nariai1}, whose sign controls
the instability. This quantity represents the initial value of the
expansion of the spatial \St-factor with respect to a foliation of
homogeneous surfaces. For $H^{(0)}_*=0$, we get a generalized Nariai
solution and the spatial \St-factor stays constant in time.  If
$H^{(0)}_*$ is an arbitrarily small positive number and the direction
of time is chosen such that the spatial \So-factor is initially
expanding, then the expansion of the \St-factor increases during the
future evolution and eventually, the foliation becomes consistent with
the cosmic no-hair picture. In this case, the solution forms a smooth
future conformal boundary. If, however, $H^{(0)}_*$ is a negative
number with arbitrarily small modulus, then the spatial \St-factor
collapses and the solution forms cigar-type curvature singularity in
the future. 

It was the idea of Bousso\cite{Bousso03} to exploit this
instability of the Nariai solutions to construct inhomogeneous
cosmological black hole solutions by making the parameter $H^{(0)}_*$
spatially dependent. The expectation is that the local value of its
sign determines whether the solutions forms either black hole interior
regions or cosmologically expanding regions. In his article he
discusses the spherically symmetric case and is able to confirm the
expectation using heuristic arguments. The resulting cosmological
black holes are the Schwarzschild-de-Sitter
solutions\cite{hawking}. It was the aim of our discussion in
\rcite{beyer09:Nariai2} to study the same problem for Gowdy symmetry.
The Gowdy case is more challenging and numerical techniques are
necessary. 
In our paper, we derive a family Gowdy initial data on $\SoXSt$ close to
certain generalized Nariai solutions and evolve them numerically so
that the quantity $H^{(0)}_*$ is spatially dependent.  

To our surprise, the numerical results, obtained with a numerical code
based on \rcite{beyer08:code}, contradict our expectations
\cite{beyer09:Nariai2}. Namely, eventually the solutions ``make a
decision'' whether the spatial \St-factor expands or collapses
\textit{globally} in space and the expected local behavior is
suppressed. There appears to be a new critical solution, i.e.\ a
critical value $\mu_c$ of a parameter $\mu$ so that for $\mu<\mu_c$,
the solution collapses globally in space, and for $\mu>\mu_c$, expands
globally in space. It would be interesting to identify the critical
solution and to study whether critical phenomena, which play such an
important role\cite{Gundlach2007} for the critical collapse of black
holes, also occur here. In summary, our results give evidence that it
is not possible to construct cosmological black hole solutions by
means of small Gowdy symmetric perturbations of the Nariai solutions.
We hope to shed further light on this for instance by means of
linearization.  Moreover, we will study ``large'' perturbations, and
our preliminary results suggest that in contrast to ``small''
perturbations those show the expected local behavior associated with
cosmological black holes.


\end{document}